%% file: SmartGrids-Text-2021_lesia.tex
\documentclass[conference]{IEEEtran}
\IEEEoverridecommandlockouts

\IEEEoverridecommandlockouts
\usepackage[dvips]{graphicx}
\usepackage[cmex10]{amsmath}
\usepackage{amsfonts,amssymb,bm}
\usepackage[footnotesize]{subfigure}
\usepackage{fixltx2e}
\usepackage{dblfloatfix}
\usepackage{rotating}
\usepackage{multirow}
\usepackage{url}
\usepackage{cite}
\usepackage[linesnumbered,lined, ruled]{algorithm2e}

\usepackage{cite}
\usepackage{setspace}
\usepackage{footnote}
\usepackage[T1]{fontenc}
\usepackage{ae,aecompl}
\usepackage{epsfig}
\usepackage{multicol}
\usepackage{multirow}
\usepackage{subfigure}
\usepackage{times}
\usepackage{color}

\usepackage{colortbl}
\usepackage[official]{eurosym}

\SetCommentSty{mycommfont}
\newlength{\commentWidth}
\SetKwRepeat{Repeat}{repeat}{until}
\SetKwRepeat{Forever}{repeat}{forever}
\SetKwRepeat{On}{on}{end}
\SetNlSty{bfseries}{\color{black}}{}

\newtheorem{remark}{Remark}

\newtheorem{definition}{Definition}[section]

\newcommand{\eg}[0]{\textit{e.g.},~}

\begin{document}

\vspace{-10pt}
\title{Game-Theoretic Energy Source Allocation Mechanism in Smart-Grids}

\vspace{-5pt}
\author{
\IEEEauthorblockN{Eleni Stai$^1$, Evangelia Kokolaki$^2$, Lesia Mitridati$^1$, Petros Tatoulis$^1$, Ioannis Stavrakakis$^3$, Gabriela Hug$^1$}\\ \vspace{-12pt}
\IEEEauthorblockA{$^1$ EEH - Power Systems Laboratory, ETH Z\"urich, Physikstrasse 3, 8092 Z\"urich, Switzerland\\
$^2$ Hellenic Ministry of Environment and Energy, Mesogeion 119, 11526 Athens, Greece\\
$^3$ Department of Informatics and Telecommunications, National \& Kapodistrian University of Athens, 15784 Athens, Greece\\
elstai@ethz.ch, e.kokolaki@prv.ypeka.gr, lmitridati@ethz.ch, petrost@student.ethz.ch, ioannis@di.uoa.gr,  ghug@ethz.ch
}
\thanks{E. Kokolaki’s work was carried out while she was with the National \& Kapodistrian University of Athens.}
}

\maketitle
\thispagestyle{empty}

\begin{abstract}
This work studies the decentralized and uncoordinated energy source selection problem for smart-grid consumers with heterogeneous energy profiles and risk attitudes: they compete for a limited amount of renewable energy in their local community, at the risk of paying a higher cost if that energy is not enough to supply all such demand. We model this problem as a non-cooperative game and study the existence of mixed-strategy Nash equilibria (NE) under the proportional allocation policy employed when the total demand for renewable energy exceeds the available one. We derive under NE closed-form expressions for  the resulting total renewable energy demand and social cost under varying consumer profiles, energy costs and availability. The analysis also provides useful guidelines as to what consumers should do (compete or not) based on their risk attitude or if they should be more risk-taking, under certain conditions. Finally, we study numerically the efficiency of this decentralized scheme compared to a centralized one via the price-of-anarchy metric.
\end{abstract}

\input{introduction}
\input{game_lesia}
\input{uncoordinated_short}

\input{evaluations}
\input{conclusions}


%
\bibliographystyle{IEEEtran}
\bibliography{references}
\input{appendix}
\clearpage

\end{document}

%% file: introduction.tex
\vspace{-0.04in}
\section{Introduction}\label{sec:intro}
\vspace{-0.04in}

Modern power systems are characterized by an increasing number of distributed renewable energy sources (RESs) that are often locally owned by active and environmentally conscious consumers. Thus, grids are driven towards a consumer-centered setting, which necessitates a paradigm shift in their operation and control. 
In this context, designing appropriate demand response programs (DRPs), which promote the adaptation of the energy consumption by self-interested consumers themselves and energy sharing within a community, has become a central question \cite{sousa2019peer}.

Several existing works in the literature on DRPs focus on the direct load control by a profit-maximizing utility and on the community-based energy sharing \cite{pinson2014benefits, moret2018negotiation,mitridati2021design}. However, these \textit{centralized} and \textit{coordinated} approaches require extensive communication among the agents, or the existence of a central coordinator (utility or community market operator), which may raise fairness \cite{vuppala2011incorporating}, privacy \cite{lisovich2010inferring} and scalability \cite{qiu2021scalable} issues. These issues may hinder the large-scale development of DRPs and should be accounted for when designing future mechanisms for demand response.
To circumvent these issues, this work focuses on the design and analysis of a \textit{decentralized} and \textit{uncoordinated} DRP for consumers in a local energy community. In the absence of a central coordinator or direct communication among consumers, the consumers decide independently how to adapt their consumption. While such a mechanism may not lead to a socially-optimal energy dispatch, its scalability and fairness properties as well as its privacy awareness render it desirable compared to traditional centralized and coordinated approaches for DRPs \cite{maharjan16,jacquot18}.

Further works in the literature have focused on the design of pricing policies to incentivize consumers to adapt their loads in a decentralized and uncoordinated manner, using non-cooperative game theoretical tools 
\cite{deng15,akbari2020concept,mitridati2021design}. 
The dynamic pricing policy proposed in \cite{Caron10} incentivizes consumers towards developing a load profile that is more convenient to be supplied by the providers. 
The distributed load management mechanism of \cite{ibars2010distributed} incentivizes consumers to shift peak consumption. Moreover, the work in \cite{jacquot18} thoroughly analyzes DRPs with an hourly billing policy. 
However, these works focus on competition among consumers across time for the capacity of a single energy source provided by a central utility, rather than selection among multiple energy sources.

In contrast with the aforementioned literature, in this paper, we are interested in studying the energy source selection problem of consumers in a local community, who independently decide which resource to consume, namely (i) locally-produced RESs generation, which is expected to cost less, is environmentally friendly and assists the economy of the local community, or (ii) centralized conventional generation, with a fixed price depending on the time of the day, based on their preferences (i.e., energy needs and attitude towards risk). Opting for RESs entails the risk of incurring a high consumer cost if the aggregate demand for RESs exceeds its available capacity. In this case, the available RES capacity is allocated among consumers based on a given \textit{allocation policy}, and the excess demand is covered by high-priced peak-load generation. Alternatively, consumers may choose to engage their loads during night when their demand will be covered by a medium-priced base load generation. In essence, each consumer faces the dilemma of competing or not for a limited inexpensive but risky resource: if they compete and are successful they incur a low resource cost; if they compete and fail, they incur a high cost; if they decide not to compete, they incur a medium cost. This general resource selection problem formulation with the ternary cost structure can model a wealth of resource selection cases, such as in the case of parking resources \cite{kokolaki2013}.

The proposed decentralized and uncoordinated DRP is of particular interest in the context of power systems with an increasing presence of local RESs, and growing environmental concerns on the consumers' side. Indeed, various studies have shown the operational, economic, and environmental benefits of the local integration and management of RESs for the grid and the consumers, such as reduced CO$_2$ emissions, operating costs, uncertainty, and congestion \cite{hatziargyriou2009quantification}. Also, recent studies suggest that the local consumption of RESs and direct involvement of consumers at the community level can facilitate the social acceptance of RESs \cite{von2018distributed,azarova2019designing}.

In order to assess the socio-economic feasibility of this decentralized and uncoordinated DRP with self-interested consumers, it is fundamental to choose an appropriate allocation policy. Various works in the literature have focused on the analysis of allocation policies for a limited resource among multiple players in a wide range of applications. First, an adequate allocation policy should aim at \textit{efficiently} incentivizing consumers to adapt their flexible demand to avoid congestion. Second, to facilitate social-acceptance and consumers engagement an adequate allocation policy should satisfy a notion of \textit{fairness}.  
In particular, the work in \cite{bertsimas2011price} showed that proportional fairness, may provide higher efficiency and a lower "cost of fairness" than other axiomatically justified notions of fairness. For this reason, the proportional allocation (PA) policy is of particular interest in our DRP and it is applied in this paper. However, contrary to our work, the work in \cite{bertsimas2011price} does not investigate agents with heterogeneous preferences, such as attitude towards risk, which limits its application to realistic local energy communities. To the best of our knowledge, no application of this notion of fairness has been proposed to design decentralized and uncoordinated DRPs in local energy communities.

Finally, when developing decentralized DRPs it is important to assess the efficiency loss compared to a centralized DRP approach. The authors in \cite{ma2011decentralized} have shown that for an infinite population of consumers with identical preferences, the NE of a distributed DRP is efficient. However, these assumptions are impractical and quite restrictive.
Further works have focused on quantifying the loss of efficiency resulting from the self-interested behavior of consumers compared to a centralized approach, using the so-called Price of Anarchy (PoA) metric, e.g., \cite{rodriguez2021value, chakraborty2017distributed}.
For instance, the authors in \cite{chakraborty2017distributed} study the distributed control of loads using a proportional allocation (PA) policy and derive a bound on the PoA.
Yet, the aforementioned papers, again, focus on the pricing/allocation of a single source of energy, rather than the allocation of multiple energy sources to consumers with heterogeneous preferences. 

Given the described research gaps, the contributions of this paper are threefold\\
$\bullet$ First, we present a novel game-theoretical formulation of the aforementioned decentralized and uncoordinated energy source selection problem for consumers in a local energy community with heterogeneous preferences, namely their energy demand and attitude towards risk. The proposed game covers a broad range of consumer types and defferable loads (e.g. electric vehicles, electrical appliances, space heating, etc.).\\
$\bullet$ Second, we study the existence of NE under the PA policy and we derive closed-form expressions for the resulting RES demand
and social cost under NE. To the best of our knowledge, this paper is the first to propose, thoroughly analyze and evaluate a game-theoretic framework for the distributed, uncoordinated energy source selection problem in local energy communities with differently-priced resources and heterogeneous consumers preferences.\\
$\bullet$ Third, we numerically assess the efficiency of the distributed, uncoordinated energy selection game with respect to the benchmark centralized mechanism using the Price-of-Anarchy (PoA) metric. 

The rest of the paper is organized as follows. Section \ref{sec:game} introduces the game-theoretic model. Section \ref{sec:analysis} studies the existence of NE and gives related closed-form expressions. 
Section \ref{sec:evaluation} presents the numerical evaluations with emphasis on the PoA metric. Finally, Section \ref{sec:conclusion} concludes the paper.

%% file: game_lesia.tex
\vspace{-0.05in}
\section{Uncoordinated energy source selection game}\vspace{-0.05in}\label{sec:game}

\subsection{Game set up}
\vspace{-0.05in}
We consider an energy community with $N$ consumers that have access to multiple energy sources in order to serve their flexible loads. During the day, they have access to (i) a limited RES capacity $\mathcal{ER}>0$ (in energy units \eg kWh) produced locally at a low-cost price $c_{RES}$ per unit of energy, and (ii) an unlimited peak-load production from the grid at a high-cost price $c_{nonRES,d}= \gamma \cdot  c_{RES}$, with $\gamma>1$. During the night, they have access to an unlimited base-load production from the grid, with a medium-cost $c_{nonRES,n} = \beta \cdot c_{RES}$, with $\gamma>\beta>1$.

At the beginning of a given day, each self-interested consumer independently decides whether they will engage, i.e., supply, their flexible loads during the day or during the night. Once engaged, their loads cannot be interrupted or shifted to another time period. If a consumer engages its load during the day, they compete with other consumers to use the low-cost RESs but incur the risk to be allocated the high-cost peak-load production during the day. Indeed, if too many consumers compete for the local RESs and their aggregate demand exceeds the available RESs capacity, the part of their load that cannot be covered locally by RESs must be covered by the peak-load production during the day. Due to this risk the consumers may wish to engage only part of their total flexible load in the day depending on their risk aversion. The remainder of their daily flexible load is then curtailed for that day, and transferred to the following day that the game is played.

This behavior represents a broad range of loads, including shiftable appliances, such as washing machines, that do not need to run every day, as well as EVs, water heaters, and batteries, that do not need to be fully charged at the end of a given day. For instance, an EV owner would compute their minimum (inflexible) daily energy load, representing the energy needed to cover their transportation needs for the day, as well as their flexible energy load, representing the additional energy needed to fully charge their EV. Then, if they decide to compete for RESs, they may be willing to engage only part of their flexible load to mitigate the risk of paying for the high-priced peak-load production. At the beginning of the following day that they play this game, they would update their daily inflexible and flexible loads and their risk attitude based on their new state-of-charge and transportation needs.

In the following, we provide the mathematical definition of this uncoordinated \emph{Energy Source Selection Game (ESSG)}.

\begin{definition}\label{def:energy_source_game}
An \emph{Energy Source Selection Game} is a tuple
\\$\Gamma=(\mathcal{N}, \mathcal{ER},  \{A_{i}\}_{i\in\mathcal{N}}, \{\vartheta_{i}\}_{i\in \mathcal{N}}, \mathbf{r}, \{v_{A_{i},\vartheta_{i}}\}_{i\in \mathcal{N}})$, where:\\
$\bullet$ $\mathcal{N}=\{1,...,N\}$, is the set of energy consumers. 
\\
$\bullet$  $\mathcal{ER}>0$ is the limited RES capacity in energy units.
\\
$\bullet$ $A_i$ is the action of player $i$ taking values in the set of potential pure strategies $\mathcal{A}=\{RES,nonRES\}$ (which is the same for all consumers). $\mathcal{A}$ consists of the choices to engage their loads during the day to compete for RES ($RES$) or engage their loads during the night ($nonRES$).\\
$\bullet$ $\vartheta_{i} \in \Theta=\{0,1,...,M-1\}$ ($M \leq N$) is the type of consumer $i \in \mathcal{N}$, consisting of an energy load profile $U_{\vartheta_i}$ and a \textit{risk aversion degree} $\mu_{\vartheta_i}\leq 1$. The risk aversion degree represents the aversion of consumers to risk their entire load during the day if playing $RES$. In particular, if a consumer $i \in \mathcal{N}$ plays $A_i = nonRES$, it engages its entire load profile $U_{\vartheta_i}$ at night. If a consumer plays $A_i = RES$, it engages a part of its load equal to $\mu_{\vartheta_i} U_{\vartheta_i}$ during the day, while the remaining of its load $(1-\mu_{\vartheta_i})U_{\vartheta_i}$ is curtailed or transferred to another time that the game is played. Therefore, $\mu_{\vartheta_i}=1$ represents a risk-seeking consumer $i$ who risks its entire load when playing $RES$, and $\mu_{\vartheta_i}<1$ a risk-conservative consumer.\\
$\bullet$ $\mathbf{r}=[r_0,...,r_{M-1}]^T$ is a probability distribution with $0\leq r_{\ell} \leq 1$ the probability that a consumer is of type $\ell \in \Theta$.
\\
$\bullet$ $\upsilon_{A_i,\vartheta_i}(.): \mathbb{R}^M \rightarrow \mathbb{R}$ is the cost function of a player $i$ with type $\vartheta_i$ and action $A_i$. 
\end{definition}

The ESSG is played once a day and the type of each consumer (energy load profile and risk aversion degree) may differ between consecutive games but remains fixed during a particular game. When taking decisions (on the variables $A_i, \forall i \in \mathcal{N}$) the consumers have information on (i) the price parameters ($c_{RES}, \beta, \gamma$), (ii) the available RESs capacity, $\mathcal{ER}$, (iii) the maximum possible expected aggregate demand for RESs (i.e., if everyone competes for RESs during the day),  $D^{Total} = N  \sum_{\ell \in \Theta} r_{\ell  }\mu_{\ell} U_{\ell}$. 

In this paper, we will study the game equilibria under mixed strategies. Let  $\mathbf{p}_{\boldsymbol \ell}=[p_{RES,\ell}, p_{nonRES,\ell}]^T$ be the \textit{mixed strategy} of consumers of type $\ell \in \Theta$. This is a probability distribution assigning to each action in the set $\mathcal{A}$ a likelihood of being selected. A consumer of type $\ell$ with a mixed strategy $\mathbf{p}_{\boldsymbol \ell}$ plays the game by randomly selecting an action in $\mathcal{A}$, with $p_{RES,\ell}$ being the probability of playing $RES$, and $p_{nonRES,\ell}$ the probability of playing $nonRES$. Note that a \textit{pure strategy} is a special case of a mixed strategy where one action has a probability equal to 1 (and the remaining have 0). We denote with $\mathbf{p}=[\mathbf{p_0}^T;\mathbf{p_1}^T;...;\mathbf{p_{M-1}}^T]$ the vector of mixed strategies for all consumer types.


In the remainder of the paper, we set the energy load profile of consumer $i$ of type $\vartheta_i$ as $U_{\vartheta_i}=\epsilon_{\vartheta_i} E_{\vartheta_i}$ and the risk aversion degree as $\mu_{\theta_i}=\frac{1}{\epsilon_{\vartheta_i}}$, with $\epsilon_{\vartheta_i}\geq 1$. Then, $ E_{\vartheta_i} = \mu_{\theta_i}U_{\vartheta_i}$ represents the load that a consumer $i$ is willing to risk during the day if they play $RES$, which is called the day-time energy demand. And $U_{\vartheta_i}=\epsilon_{\vartheta_i} E_{\vartheta_i}$ represents the entire load that the consumer engages during the night if they play $nonRES$. Also, $\epsilon_{\vartheta_i}=1$ represents a risk-seeking consumer $i$, and $\epsilon_{\vartheta_i}>1$ a risk-conservative consumer. From now on we refer to $\epsilon_{\vartheta_i}$ as the inverse risk aversion degree. Finally, we assume without loss of generality that $E_0\leq E_1 \leq ...\leq E_{M-1}$.

\vspace{-0.08in}
\subsection{Energy allocation policy}\label{sec:policies}
\vspace{-0.05in}
The expected aggregate demand for RESs is expressed as

\vspace{-0.1in}
\begin{small}
\begin{equation}
    D (\mathbf{p}) = N\sum_{\ell\in \Theta} r_{\ell }~p_{RES,\ell}~E_{\ell}. \label{eq:demand}
\end{equation}
\end{small}
If this expected aggregate demand exceeds the available RESs capacity $\mathcal{ER}$, and since the individual consumer loads cannot be interrupted or shifted, the excess demand must be covered by the peak-load production during the day. In order to allocate the available RES capacity fairly among consumers, this paper examines the PA policy. Under PA, the share of RESs received by a consumer $i$ of type $ \vartheta_i \in \Theta$ that plays $RES$, $rse_{\vartheta_i}^{PA}(\mathbf{p})$, is proportional to its demand, and depends on the strategies of all consumers, such that

\vspace{-0.15in}
\begin{small}
\begin{eqnarray}
rse_{\vartheta_i}^{PA}(\mathbf{p}) &=& \frac{E_{\vartheta_i}}{\max(\mathcal{ER} , D (\mathbf{p}))}\mathcal{ER}.
\label{eq:prop_alloc_energy}
\end{eqnarray}
\end{small}
\vspace{-0.3in}
\subsection{Cost function}\label{sec:costs} 
\vspace{-0.05in}
The cost function $\upsilon_{RES,\vartheta_i}(.)$ for a consumer $i$ of type $\vartheta_i$ that plays $RES$ depends on the strategies of all consumers and the chosen allocation policy, and can be expressed as

\vspace{-0.15in}
\begin{small}
\begin{align}
& \upsilon_{RES,\vartheta_i}(\mathbf{p}) =
rse_{\vartheta_i}^{PA}(\mathbf{p}) \cdot c_{RES} + (E_{\vartheta_i}-rse_{\vartheta_i}^{PA}(\mathbf{p}) ) \cdot c_{nonRES,d}.
\label{eq:RES_cost}
\end{align}
\end{small}
\noindent Note that the load that is not enganged, i.e., $(1-\mu_{\vartheta_i})U_{\vartheta_i}$ does not incur any cost. Moreover, the cost $\upsilon_{nonRES,\vartheta_i}(.)$ for a consumer $i$ of type $\vartheta_i$ that chooses $nonRES$ is 

\vspace{-0.1in}
\begin{small}
\begin{equation}
\upsilon_{nonRES,\vartheta_i}=\epsilon_{\vartheta_i} \cdot E_{\vartheta_i} \cdot c_{nonRES,n},
\label{eq:nonRES_cost2}
\end{equation}
\end{small}
and solely depends on that consumer's strategy.

%% file: uncoordinated_short.tex
\vspace{-0.05in}
\section{Game analysis} \label{sec:analysis}
\vspace{-0.05in}
Here, we study the conditions on the parameter values for the existence of dominant strategies and mixed-strategy NE under PA. For this study, we distinguish cases with respect to the RES capacity, the risk aversion degree values and the day-time energy demand levels. 
\vspace{-0.05in}
\subsection{Case $1$}\vspace{-0.025in} \emph{The RES capacity, $\mathcal{ER}$, exceeds the quantity $D^{Total}$.} As the consumers have knowledge of $\mathcal{ER}$ and $D^{Total}$, it is straightforward to show that the dominant strategy for all consumers is to play $RES$ and the aggregate demand for RESs is equal to $D^{Total}$.

In all the remaining cases, we assume that $\mathcal{ER}< D^{Total}$. 
\vspace{-0.05in}
\subsection{Case $2$}\vspace{-0.025in} \emph{The inverse risk aversion degrees for all consumers' types satisfy $1\leq \epsilon_{\ell}<\gamma/\beta$, $\forall \ell \in \Theta$.}
We distinguish the following sub-cases with respect to the day-time energy demand levels. 

\subsubsection{Sub-case $2(a)$} 
\emph{The day-time energy demand levels satisfy  $E_{\ell } \leq \mathcal{ER}\frac{(\gamma-1)}{(\gamma-\epsilon_{\ell }\beta)}$,  for all $\ell \in \Theta$.} Then, a mixed-strategy NE with the PA policy exists if and only if for every pair of consumers, $i,j$ with day-time energy demand levels $E_{\vartheta_i}$, $E_{\vartheta_j}$, respectively, it holds that

\vspace{-0.15in}
\small
\begin{eqnarray}\label{eq:relation_E_0_E_1_pa_ne_extra_demand}
\mathcal{ER}\frac{(\gamma-1)}{(\gamma-\epsilon_{\vartheta_i }\beta)}-E_{\vartheta_i} &=& \mathcal{ER}\frac{(\gamma-1)}{(\gamma-\epsilon_{\vartheta_j }\beta)}-E_{\vartheta_j}.
 \end{eqnarray}
\normalsize
Furthermore, at NE, the expected aggregate demand for RES can be expressed, for any $\ell \in \Theta$, as:

\footnotesize
\begin{align}
& D(\mathbf{p}^{NE})= \min\Bigl\{D^{Total}, \max\Bigl\{\left[\mathcal{ER}\frac{(\gamma-1)}{(\gamma-\epsilon_{\ell}\beta)}-E_{\ell}\right]\frac{N}{(N-1)},0\Bigr\}\Bigr\},
    \label{eq:demand1}
\end{align}
\normalsize
\noindent with $\mathbf{p}^{NE}$ the NE mixed strategies. The proofs are in Appendix \ref{sec:subcase2a}.
\begin{remark}\label{rem:risk_seeking}If all consumers are risk-seeking (i.e., $\epsilon_{\vartheta_i}= 1, \forall i \in \mathcal{N}$), a NE can exist only if  $E_{0}=E_{1}=...=E_{M-1}$.
\end{remark}

\begin{remark} \label{rem:risk_degrees_relation}
Note that condition \eqref{eq:relation_E_0_E_1_pa_ne_extra_demand} can hold, and therefore a NE can exist, only if $\epsilon_0\leq \epsilon_1 \leq..\leq \epsilon_{M-1}$. Since by assumption, $E_0\leq E_1 \leq ...\leq E_{M-1}$, this means that consumers with lower day-time energy demand levels should be less risk-averse than those with higher ones.
\end{remark}

\subsubsection{Sub-case $2(b)$} 
\emph{The day-time energy demand levels satisfy $E_{\ell} > \mathcal{ER}\frac{(\gamma-1)}{(\gamma-\epsilon_{\ell}\beta)}$  for all $\ell \in \Theta$.} In this case, the dominant strategy for all consumers is to play $nonRES$ and cover their entire energy profiles ($U_{\ell}$) at night. Thus, $D^{Total}=0$.

The proof is in Appendix \ref{sec:subcase2b}.

\subsubsection{Sub-case $2(c)$}
\emph{There exist two distinct subsets of consumers' types, $\Sigma_1 , \Sigma_2 \subset \Theta$, where $\{E_{\ell} > \mathcal{ER}\frac{(\gamma-1)}{(\gamma-\epsilon_{\ell}\beta)}, ~\forall \ell \in \Sigma_1\}$ and $\{E_{\ell} \leq \mathcal{ER}\frac{(\gamma-1)}{(\gamma-\epsilon_{\ell}\beta)}, ~ \forall \ell \in \Sigma_2\}$.} For consumers in the set $\Sigma_1$, the dominant strategy is to play $nonRES$ and to cover their entire energy profiles ($U_{\ell}$) during the night.
For consumers in the set $\Sigma_2$, the mixed strategy NE is determined under the condition of \eqref{eq:relation_E_0_E_1_pa_ne_extra_demand}. The aggregate demand for RES is given by \eqref{eq:demand1} if replacing $D^{Total }$ with $D^{Total}_{\Sigma_2} = N  \sum_{\tilde{\ell } \in \Sigma_2} r_{\tilde{\ell }  }E_{\tilde{\ell } }$.
\vspace{-0.05in}
\subsection{Case $3$}\vspace{-0.025in} \emph{The inverse risk aversion degrees satisfy $\epsilon_{\ell} \geq \gamma/\beta$, $\forall \ell \in \Theta$.} In this case, the dominant strategy for all consumers is to play $RES$.

The proof is in Appendix \ref{sec:subcase3}.
\vspace{-0.05in}
\subsection{Case $4$}\vspace{-0.025in} \emph{There exist two distinct subsets of consumers' types, $\Sigma_1$, $\Sigma_2 \subset \Theta$, where $\Bigl\{\epsilon_{\ell} \geq \gamma/\beta, ~ \forall \ell \in \Sigma_1\Bigr\}$, and $\Bigl\{\epsilon_{\ell} < \gamma/\beta,~\forall \ell \in \Sigma_2\Bigr\}$.} For consumers in the set $\Sigma_1$, the dominant strategy is to compete for RESs, which is shown by analogy with Case $3$.
For consumers in the set $\Sigma_2$, the conditions derived in Case $2$ can be directly extended to the types in $\Sigma_2$, considering that all consumers in $\Sigma_1$ play $RES$. To do so, we should consider that the available RES capacity for consumers in $\Sigma_2$ is equal to $\mathcal{ER}$ minus the aggregate RESs demand of $\Sigma_1$. For the sake of concision we do not detail these conditions.

%% file: evaluations.tex
\vspace{-0.05in}
\section{Evaluations}
\label{sec:evaluation}
\vspace{-0.05in}
In this section, we numerically evaluate the proposed decentralized, uncoordinated DRP for different parameter values and study the strategies, the demand and the social cost at NE. Any existing mixed-strategy NE, $\mathbf{p}^{NE}$, are obtained so as to satisfy \eqref{eq:condition_PA_NE}. In addition, we perform comparisons with a centralized mechanism that allocates the consumers' loads between day and night so as to minimize the expected social cost. The expected social cost for allocation $\mathbf{p}$ is defined as

\vspace{-0.15in}
\begin{small}
\begin{align}
C(\mathbf{p}) &=  \min(\mathcal{ER}, D(\mathbf{p})) c_{RES} +  \max(0,D(\mathbf{p})-\mathcal{ER}) \nonumber \\& \cdot c_{nonRES,d}+N \left[ \sum_{\ell \in \Theta} r_{\ell }p_{nonRES,\ell}\epsilon_{\ell }E_{\ell}\right] c_{nonRES,n},
\label{eq:social_cost_pa_extra_demand}
\end{align}
\end{small}

\noindent where the first two summands refer to the consumers who played $RES$ and the third one to those who played $nonRES$. 

Now, the (in)efficiency of NE strategies is quantified by the PoA \cite{Koutsoupias09}, which is expressed as the ratio of the worst case social cost among all NE over the optimal-minimum social cost of the centralized mechanism. 
\vspace{-0.06in}

\subsection{Two-type smart-grid}\vspace{-0.06in}
In order to quickly gain insights into the centralized mechanism and the decentralized energy selection game, we first study a simple smart-grid with $N=500$ consumers of two types $l \in \{0,1\}$. The RES capacity is $\mathcal{ER}=10$ kW, and its cost is $c_{RES}=0.3$ \euro/kWh. 

\subsubsection{Competing Probabilities}
\begin{figure}[b]\vspace{-0.2in}
    \centering
    \includegraphics[scale=.2]{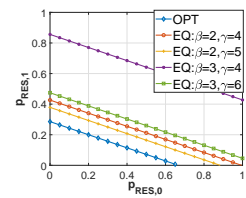}
    \caption{Optimal and NE competing probabilities.}\vspace{-0.2in}
    \label{fig:comp_prob}
\end{figure}

We first consider solely risk-seeking consumers, with fixed day-time energy demand values $E_0=E_1=100$ kW (where the equality is required for a NE to exist), and types' distribution $r_0=0.3$, $r_1=0.7$.
Fig. \ref{fig:comp_prob} illustrates the optimal solution to the centralized problem (OPT) and the NE solution of the energy source selection game (EQ) for different values of prices $\beta$ and $\gamma$. 
As expected, there exist infinitely many optimal and equilibrium solutions, which consist of asymmetric probabilities for the two types. Consumers show greater willingness to compete for RES in absence of coordination. In addition, for the same night-time price ($\beta$), consumers are competing with higher probabilities for lower day-time price ($\gamma$). On the other hand, for equal values of $\gamma$, a higher $\beta$ leads to higher competing probabilities. Finally, for the same ratio of $\gamma / \beta$, the competing probabilities are higher for higher price levels $\beta$ and $\gamma$.

\subsubsection{Social Cost \& PoA} 
\begin{figure}[t]
     \centering
     \subfigure[Social Cost (in eurocents). \label{fig:3a}]{
         \includegraphics[width=0.22\textwidth]{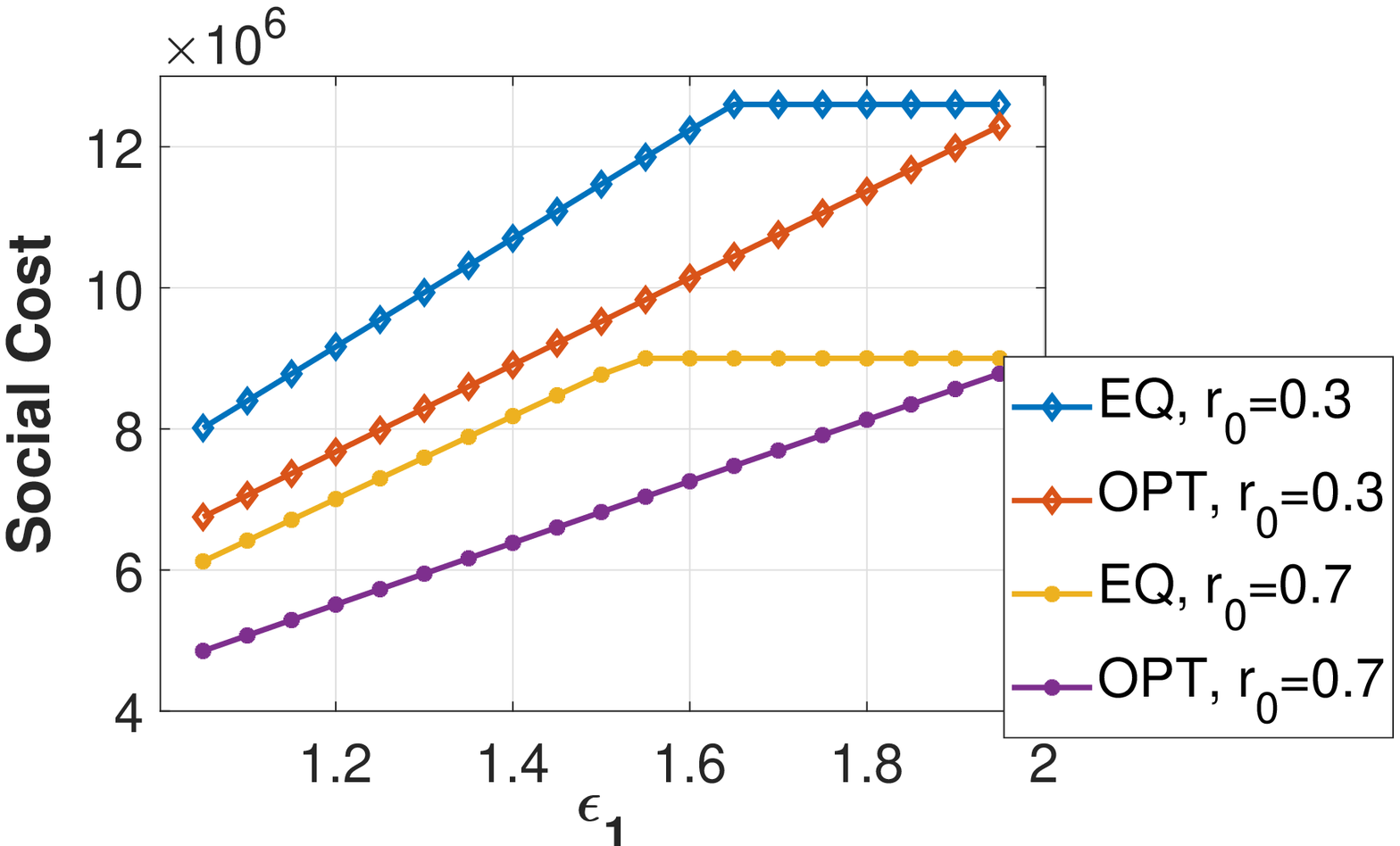}}
  \subfigure[PoA. \label{fig:3b}]{
         \includegraphics[width=0.205\textwidth]{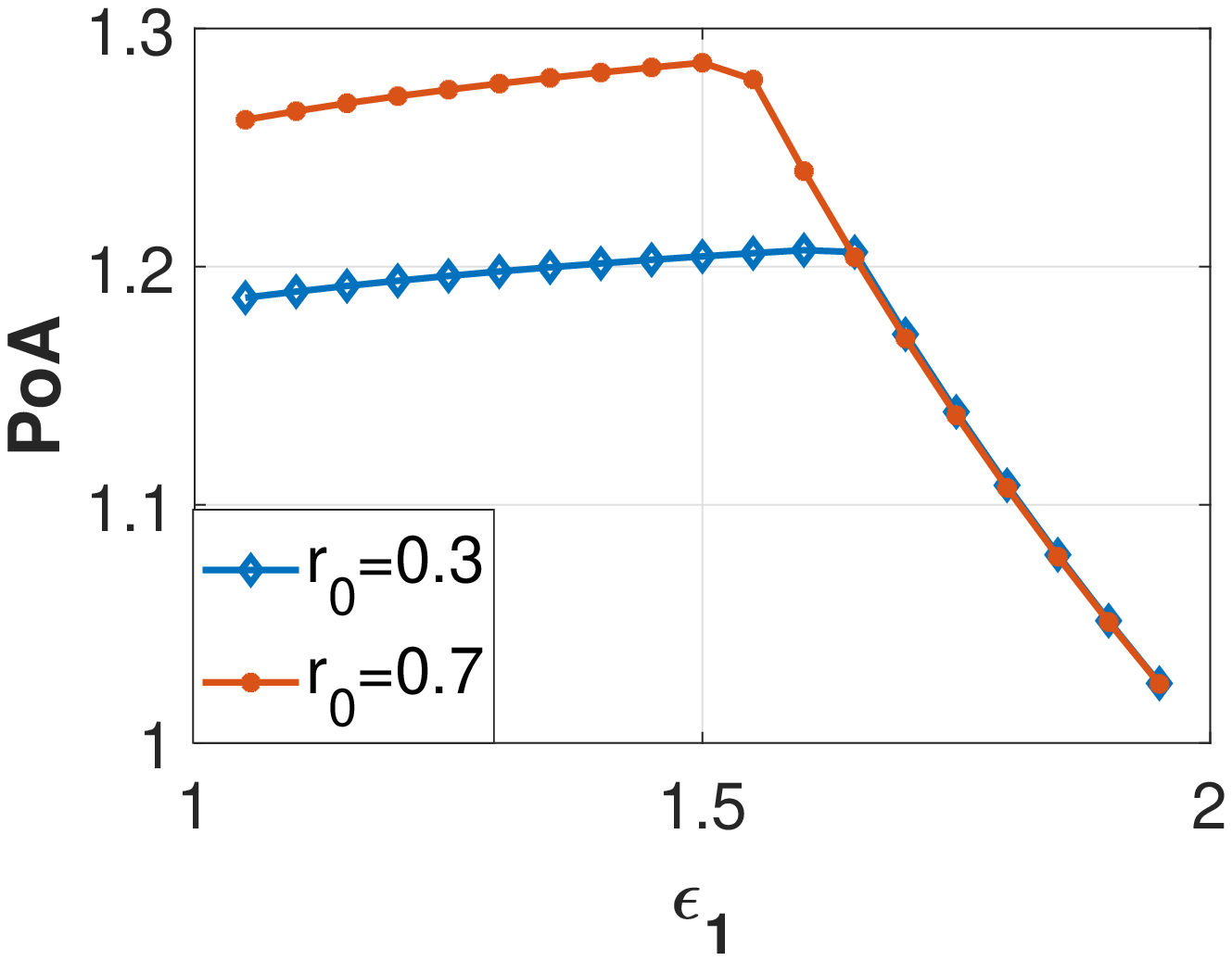}}
        \caption{Social cost and PoA for $\epsilon_0 \leq \gamma/\beta$ under the PA rule.}
        \label{fig:3ab}
        \vspace{-0.2in}
\end{figure}

\begin{figure}[t]
     \centering
     \subfigure[Social Cost (in eurocents). \label{fig:3c}]{
         \includegraphics[width=0.2\textwidth]{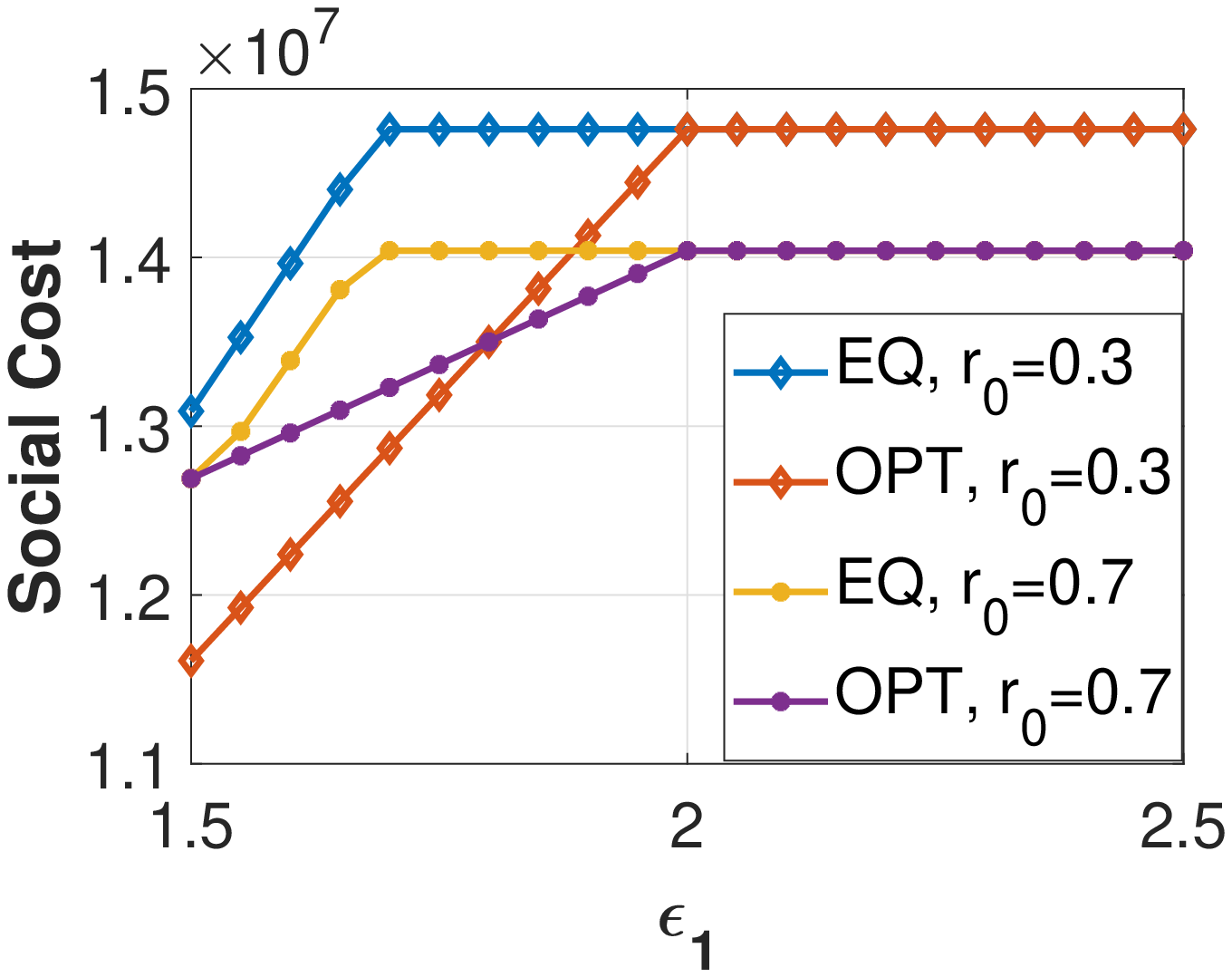}}
  \subfigure[PoA. \label{fig:3d}]{
         \includegraphics[width=0.2\textwidth]{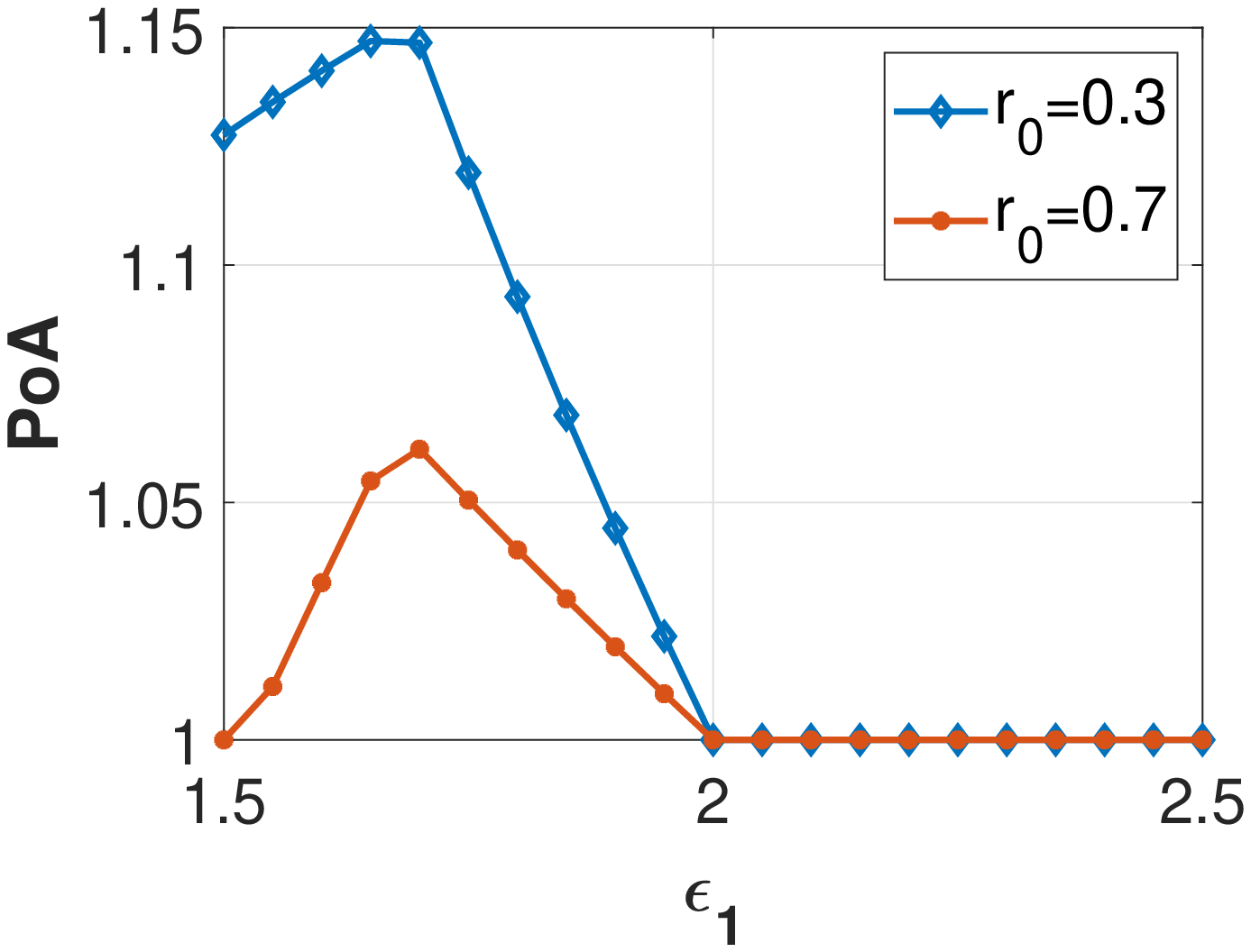}}
        \caption{Social cost and PoA for $\epsilon_0 \geq \gamma/\beta$ under the PA rule.}
        \label{fig:3cd}
        \vspace{-0.25in}
\end{figure}

We now consider fixed prices $\beta=5$, $\gamma=10$, with day-time energy demands $E_0=100$, $E_1=180$ in Fig. \ref{fig:3ab} and $E_0=100$, $E_1=200$ in Fig. \ref{fig:3cd}.
Figs. \ref{fig:3ab} and \ref{fig:3cd} illustrate the social costs and PoA for varying values of inverse risk aversion degree $\epsilon_1$ and consumer type distributions. The inverse risk aversion degree $\epsilon_0$ is determined based on $\epsilon_1$ so that the relationship \eqref{eq:relation_E_0_E_1_pa_ne_extra_demand} holds. With these values: (i) in Fig. \ref{fig:3ab} it always holds that $\epsilon_1, \epsilon_0 \leq \gamma/\beta$ and Case 2 of the analysis applies, and (ii) in Fig. \ref{fig:3cd} it either  holds that $\epsilon_1 \leq \gamma/\beta$, $\epsilon_0 \geq \gamma/\beta$ where Case 4 applies or $\epsilon_1, \epsilon_0 \geq \gamma/\beta$ in which case a dominant strategy exists by Case 3.

From Fig. \ref{fig:3a}, we observe that the social cost for the centralized optimal allocation is increasing linearly with $\epsilon_1$. This happens because under the PA policy, at the optimal solution, the demand is equal to the RES capacity $\mathcal{ER}$ and thus the social cost is driven by the night-time cost (see second line of \eqref{eq:social_cost_pa_extra_demand}) that is linear in $\epsilon_1$. For the uncoordinated strategy, the NE social cost is also increasing linearly and is always greater than the optimal social cost, which indicates that consumers tend to over-compete for RES. As $\epsilon_1$ increases, consumers of type $1$ compete more for RES since, otherwise, their night-time cost would increase. At a certain point onward they compete with probability $p_{RES,1}^{NE} = 1$ and the social cost stabilizes. Furthermore, from Fig. \ref{fig:3a} we can see that the social cost is higher for $r_0=0.3$, since in this case most of the consumers have greater day-time energy demand ($E_1>E_0$). 

Fig. \ref{fig:3b} shows the PoA for the same case. We can identify two regimes: one where PoA increases (for lower $\epsilon_1$) and a second, where the PoA decreases (higher $\epsilon_1$). In the first regime, the value of PoA is relatively large, indicating an inefficiency of around $20-30\%$. This is because the NE social cost increases more than the optimal cost with $\epsilon_1$, since consumers are over-competing for RESs and are thus paying for high-priced day-time nonRES. 
In the second regime, PoA decreases because the social cost for the equilibrium strategy flattens. However, we see that PoA falls at the same rate, regardless of the distribution of consumers. 

Similar observations apply also for Fig. \ref{fig:3cd}. In particular, for $\epsilon_1 < \gamma/\beta=2$, the trends look the same as in Fig. \ref{fig:3b}, with the difference that the PoA obtains lower values, indicating higher efficiency. For $\epsilon_1 > \gamma/\beta$, we have $100\%$ efficiency since all consumers have a dominant strategy to compete for RES, which coincides with the optimal allocation. 
\vspace{-0.07in}
\subsection{Residential Smart Grid}
\vspace{-0.07in}
Here, we study a smart grid with 5 consumer types and parameters as shown in Table \ref{tab:residential}. The type distribution and the day-time energy demand levels are selected so as to resemble the figure of the European households \cite{enerdata}.
\begin{table}[t]
    \vspace{-0.15in}
\footnotesize
    \centering
\caption{Game parameters for residential smart-grid.}\vspace{-0.1in}
    \begin{tabular}{|c||c|c|c|c|c|}
        \hline
        Type $\ell$ & 0 & 1 & 2 & 3 & 4 \\
        \hline 
        $E_\ell$ & 2 & 3 & 5 & 10 & 15 \\
        \hline
        $r_\ell$ & 0.20 & 0.40 & 0.30 & 0.07 & 0.03 \\
        \hline
    \end{tabular}
    \vspace{-0.1in}
    \label{tab:residential}
\end{table}
From Table \ref{tab:residential}, we can observe that most households are moderately energy efficient (types $1$ and $2$), combined with a significant amount of highly efficient households (type $0$) and a few fairly inefficient ones (types $3$ and $4$). Consumers with type $0$ are assumed to be risk-seeking ($\epsilon_0=1$) and the inverse risk aversion degrees of all other types are then determined by  \eqref{eq:relation_E_0_E_1_pa_ne_extra_demand}. 
However, the derived inverse risk aversion degrees of all other types are very close to $1$. Also we set $N=1000$, $c_{RES}=1$ \euro/kWh, $\beta=2$ and $\gamma=3$.

\begin{figure}[t] 
     \centering
     \subfigure[Social Cost (in eurocents). \label{fig:4a}]{
         \includegraphics[width=0.2\textwidth]{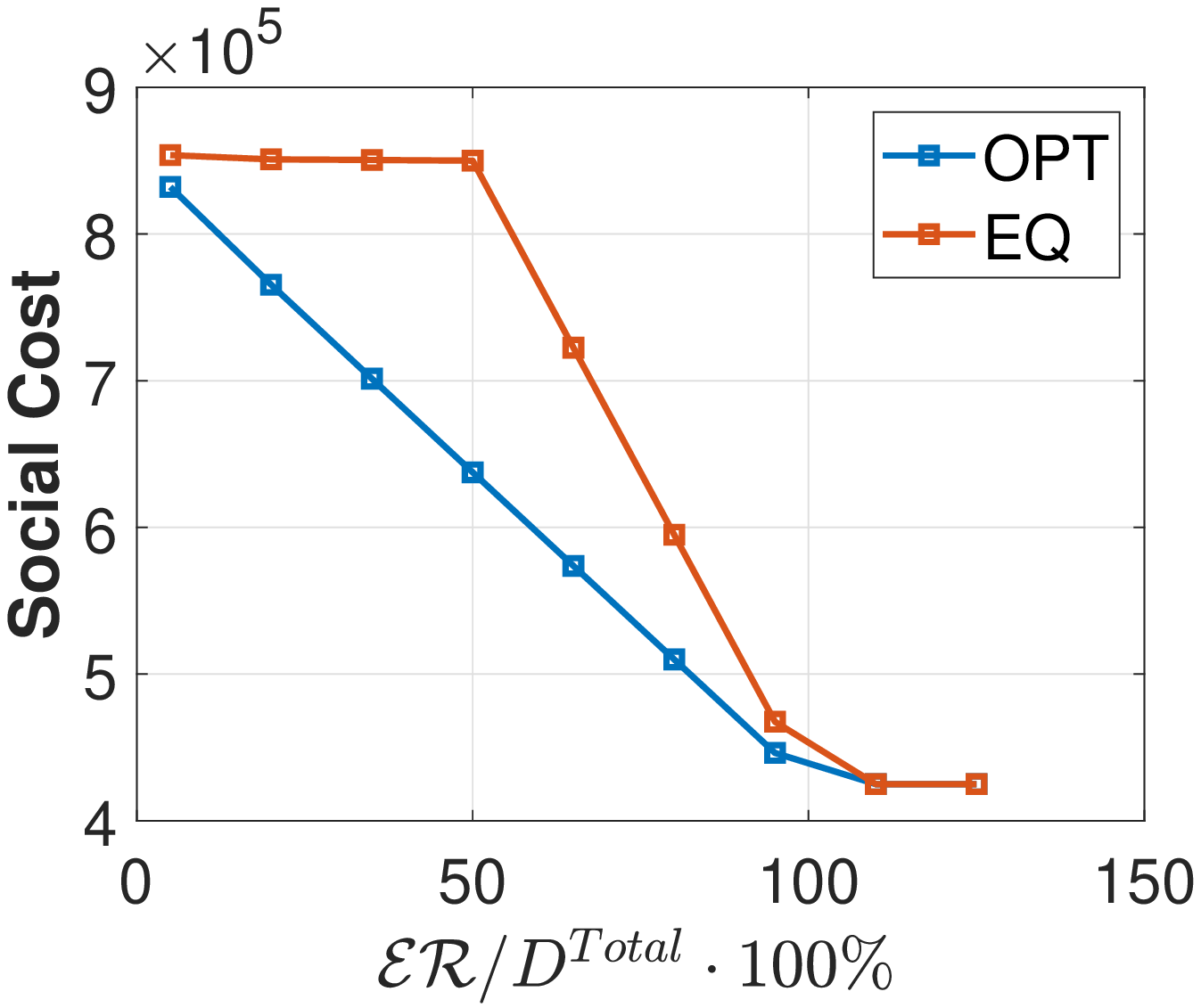}}
  \subfigure[PoA. \label{fig:4b}]{
         \includegraphics[width=0.2\textwidth]{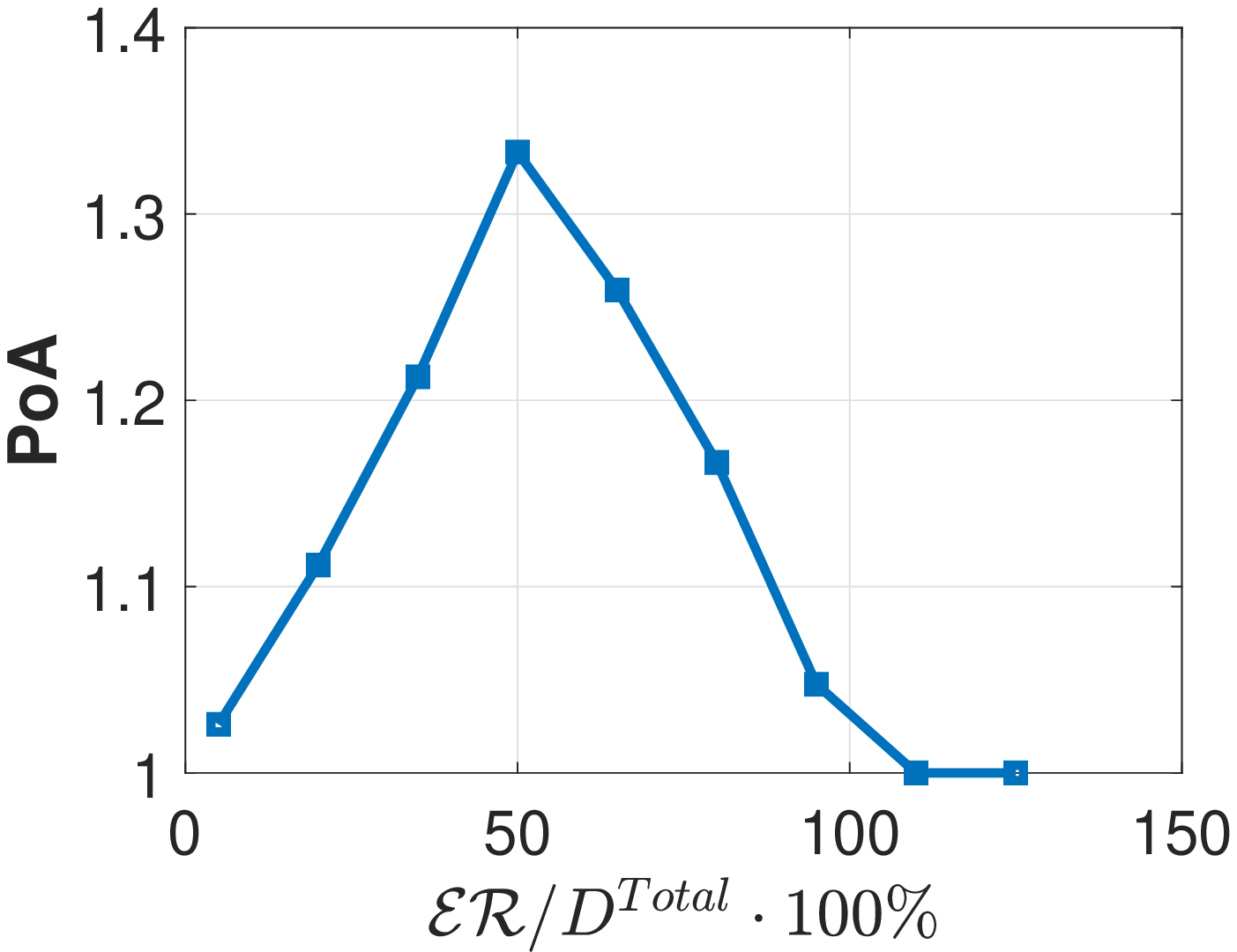}}
        \caption{Social cost and PoA under PA rule for residential grid.}\vspace{-0.25in}
        \label{fig:sc_PoA_PA}
\end{figure}

We can compute $D^{Total} = 4250$. Then, Fig. \ref{fig:sc_PoA_PA} shows the social cost and the PoA with increasing $\mathcal{ER}$ that ranges from $5\%$ to $125\%$ of $D^{Total}$. First, we observe that the optimal social cost for the centralized approach decreases linearly with $\mathcal{ER}$. This is because the optimal allocation should equalize the RES demand and capacity, while minimizing the night-time cost. Since here all risk aversion degrees are equal or close to 1, the cost at night can be approximated as {\small$N \sum_{{\ell} \in \Theta} r_{{\ell} } \cdot p_{nonRES,{\ell}} \cdot \epsilon_{{\ell} } \cdot E_{{\ell}} \cdot c_{nonRES,n}\approx N \sum_{{\ell} \in \Theta} r_{{\ell} } \cdot p_{nonRES,{\ell}} \cdot E_{{\ell}} \cdot c_{nonRES,n}\approx$ $N \sum_{{\ell} \in \Theta} r_{{\ell} } \cdot  E_{{\ell}} \cdot c_{nonRES,n} - \mathcal{ER}\cdot c_{nonRES,n} $}, which does not depend on probabilities and is linearly decreasing with $\mathcal{ER}$. Using the second in row approximation, we point out one more property of the optimal allocation: the night-time cost is mostly affected by the day-time energy demands and its minimization results in "big players" competing for RESs at the expense of smaller ones. This is indeed observed in our evaluations where consumers with lower day-time energy demand compete for RES with non-zero probability only if all consumers with higher day-time energy demand compete for RES with probability 1 and RES capacity is still available. 

On the other hand, for the uncoordinated mechanism the social cost decreases linearly with $\mathcal{ER}$, only when $\mathcal{ER}\in [0.5 D^{Total}, D^{Total}]$. The social cost barely reacts to the initial increase in RES capacity due to the fact that consumers overcompete for RES. As $\mathcal{ER}$ increases so do the NE competing probabilities and thus the demand for RES. Therefore, there is extra demand, which costs the high day-time prices. When $\mathcal{ER}\in [0.5 D^{Total}, D^{Total}]$ all consumer types compete for RES and the social cost is falling because as $\mathcal{ER}$ is increasing, fewer high-priced day-time energy is required.

Fig. \ref{fig:4b} shows the PoA. The most inefficient outcome is  for $\mathcal{ER} = 0.5 \cdot D^{Total}$, which is the point that the social cost for the uncoordinated mechanism begins decreasing. This graph can provide valuable insight into how much RES capacity we should install. 
We see two zones of high efficiency, namely for low and high RES capacity. In the first zone, this is due to the small gains in cost offered by low RES capacity. In the second, the NE solution has almost converged to the optimal solution and thus social costs are optimal as well. 

%% file: conclusions.tex
\vspace{-0.05in}
\section{Conclusions}
\label{sec:conclusion}
\vspace{-0.05in}
In this paper, we introduce and thoroughly analyze the properties of a decentralized and uncoordinated energy source selection mechanism for consumers in a local energy community with with multiple energy sources through a game-theoretic framework. 
Numerical results show that, consumers tend to over-compete for low-priced RESs. This inefficiency mostly grows with decreasing price difference between energy sources and peaks at a particular risk aversion.
This work provides the basis to design regulatory changes that incentivize investment in the appropriate ratio of local RESs, and offer adequate energy grid prices to alleviate inefficiencies. 

\vspace{-0.1in}

%% file: appendix.tex
\appendices
\vspace{-0.05in}

\section{Proofs for Sub-case $2(a)$} \label{sec:subcase2a}
Recall that any mixed strategy NE $\mathbf{p^{NE}}$
must fulfill

\vspace{-0.1in}
\small
\begin{equation}\label{eq:cost_equality_mixed}
\upsilon_{RES, \ell}(\mathbf{p^{NE}})= \upsilon_{nonRES, \ell}(\mathbf{p^{NE}}), ~\forall \ell \in \Theta.
\end{equation}
\normalsize
Namely, the expected costs of each pure strategy in the support of the equilibrium mixed strategy ($\mathcal{A}$) are equal. By replacing in \eqref{eq:cost_equality_mixed} the expressions of \eqref{eq:RES_cost} and \eqref{eq:nonRES_cost2}, we obtain that at a NE the amount of RESs allocated to $\ell \in \Theta$ should satisfy

\vspace{-0.1in}
\small
\begin{equation}\label{eq:conditionEQ_extra_demand}
rse^{PA, NE}_{\ell}(\mathbf{p^{NE}}) = \frac{\gamma-\epsilon_{\ell}\beta}{\gamma-1}E_{\ell},~ \forall \ell \in \Theta.
\end{equation}
\normalsize
Thus, when combining the Energy Source Selection Game with the PA policy, the existence of a NE is under the condition

\vspace{-0.1in}
\small
\begin{equation}\label{eq:condition_PA_NE}
rse_{\ell}^{PA}(\mathbf{p}^{NE}) =rse_{\ell}^{PA,NE}(\mathbf{p}^{NE}), \forall \ell \in \Theta. \end{equation}
\normalsize

To derive the condition \eqref{eq:relation_E_0_E_1_pa_ne_extra_demand} we re-write \eqref{eq:condition_PA_NE} first with assuming that consumer $i$ with type $\vartheta_i \in \Theta$ plays $RES$ (in \eqref{eq:probrelation1}) and second with assuming that consumer $j$ with type $\vartheta_j \in \Theta \setminus \{\vartheta_i\}$ plays RES (in \eqref{eq:probrelation2}):

\vspace{-0.1in}
\begin{small}
\begin{align}
&  \mathcal{ER}\frac{(\gamma-1)}{(\gamma-\epsilon_{\vartheta_i}\beta)}-E_{\vartheta_i}= \sum_{ {\ell}\in \Theta} r_{\ell}~ (N-1)~E_{\ell}~p^{NE}_{RES,\ell},
    \label{eq:probrelation1}\\
  &  \mathcal{ER}\frac{(\gamma-1)}{(\gamma-\epsilon_{\vartheta_j}\beta)}-E_{\vartheta_j}=  \sum_{ {\ell}\in \Theta} r_{\ell} ~(N-1)~E_{\ell}~ p^{NE}_{RES,\ell}.
    \label{eq:probrelation2}  
\end{align}
\end{small}
Note that to derive (\ref{eq:probrelation1}) we consider that if consumer $i$ plays $RES$, then, $D(\mathbf{p^{NE}})$ can be expressed as $E_{\vartheta_i}+ \sum_{ {\ell}\in \Theta} r_{\ell}~ (N-1)~E_{\ell}~p^{NE}_{RES,\ell}$ for a large number of consumers. Similarly for (\ref{eq:probrelation2}). Then, since the right-hand sides of \eqref{eq:probrelation1}-\eqref{eq:probrelation2} are equal, the left-hand sides will be also equal and \eqref{eq:relation_E_0_E_1_pa_ne_extra_demand} derives. 

Next, to construct the expression of the aggregate demand for RESs \eqref{eq:demand1}, we multiplied \eqref{eq:probrelation1} with $\frac{N}{N-1}$ and used the definition of the demand for RESs in \eqref{eq:demand}.
\vspace{-0.05in}

\section{Proof for Sub-case $2(b)$} \label{sec:subcase2b} We need to show that $\upsilon_{RES, \vartheta_i}(\mathbf{p})>\upsilon_{nonRES, \vartheta_i}(\mathbf{p})$, $\forall \mathbf{p}$ and $\forall i\in \mathcal{N}$. Assume that $\vartheta_i=0$ and that the allocated energy is $E'$. Then, we have that $\upsilon_{RES, 0}(\mathbf{p})= E'  \cdot c_{RES}+(E_0-E')\cdot \gamma \cdot  c_{RES}$ and 
$\upsilon_{nonRES, 0}(\mathbf{p})= \epsilon_0 \cdot E_0 \cdot \beta \cdot c_{RES}$. The inequality $\upsilon_{RES,0}(\mathbf{p})>\upsilon_{nonRES, 0}(\mathbf{p})$ is then equivalent to the inequality $E_0 >E' \frac{(\gamma-1)}{(\gamma-\epsilon_0\beta)}$, which is true by assumption, since $E'<\mathcal{ER}$. A similar proof can be constructed for every consumer and consumer type in $\Theta$.
\vspace{-0.05in}
\section{Proof for Case $3$} \label{sec:subcase3}To show this, assume that $\ell=0$ and that the allocated energy is $E'$. Then, the inequality $\upsilon_{RES,0}(\mathbf{p})<\upsilon_{nonRES, 0}(\mathbf{p})$ is equivalent to the inequality $E_0 >E' \frac{(\gamma-1)}{(\gamma-\epsilon_0\beta)}$. This is true by assumption, since $(\gamma-\epsilon_0\beta)\leq 0$ in this case. A similar proof can be constructed for every type $\ell \in \Theta$.